# PT-symmetry in nonlinear twisted multi-core fibers

Xiao Zhang,[1] Victor A. Vysloukh,[2] Yaroslav V. Kartashov,[3,4,5] Xianfeng Chen,[1] Fangwei Ye,[1,*] and Milivoj R. Belić[6]

[1]Key Laboratory for Laser Plasma (Ministry of Education), Collaborative Innovation Center of IFSA, Department of Physics and Astronomy, Shanghai Jiao Tong University, Shanghai 200240, China
[2]Departamento de Fisica y Matematicas, Universidad de las Americas Puebla, 72820 Puebla, Mexico
[3]ICFO-Institut de Ciencies Fotoniques, The Barcelona Institute of Science and Technology, 08860 Castelldefels (Barcelona), Spain
[4]Institute of Spectroscopy, Russian Academy of Sciences, Troitsk, Moscow Region, 142190, Russia
[5]Department of Physics, University of Bath, BA2 7AY, Bath, United Kingdom
[6]Science Program, Texas A&M University at Qatar, P.O. Box 23874 Doha, Qatar
*Corresponding author: fangweiye@sjtu.edu.cn

**We address propagation of light in nonlinear twisted multi-core fibers with alternating amplifying and absorbing cores that are arranged into the $\mathcal{PT}$-symmetric structure. In this structure, the coupling strength between neighboring cores and global energy transport can be controlled not only by the nonlinearity strength, but also by gain and losses and by the fiber twisting rate. The threshold level of gain/losses, at which $\mathcal{PT}$-symmetry breaking occurs, is a non-monotonic function of the fiber twisting rate and it can be reduced nearly to zero or, instead, notably increased just by changing this rate. Nonlinearity usually leads to the monotonic reduction of the symmetry breaking threshold in such fibers.**

http://dx.doi.org/10.1364/OL.99.099999

The concept of parity-time $(\mathcal{PT})$ symmetry that was initially introduced in quantum mechanics [1], has already penetrated into many other areas of science (see [2,3] for recent reviews). Various optical realizations of the $\mathcal{PT}$-symmetric systems, such as couplers, multi-core fibers, shallow photonic lattices, and photonic crystals with inhomogeneous refractive index landscapes obeying the $\mathcal{PT}$ symmetry condition $n(\mathbf{r}) = n^*(-\mathbf{r})$, where $n(\mathbf{r})$ is the complex refractive index, were suggested. Despite the presence of gain and losses in such systems, the internal currents from amplifying to absorbing domains make it possible for the propagation of the beam without net amplification or attenuation. The most representative property of the $\mathcal{PT}$-symmetric system is the existence of the threshold level of gain/losses, above which the spectrum of the system becomes complex and the propagation of the modes is always accompanied by their amplification or attenuation [4]. The breakup of the $\mathcal{PT}$ symmetry was observed experimentally [5,6]. $\mathcal{PT}$-symmetric structures that remain invariable in the direction of light propagation have been used for demonstration of the switching, localization, and nonreciprocal soliton scattering [7-13]. At the same time, longitudinal variation of the parameters of a $\mathcal{PT}$-symmetric system substantially enrich the spectrum of the available phenomena. Such dynamic structures were used for illustration of the pseudo-$\mathcal{PT}$ symmetry [14-16], dynamic localization [17,18], mode conversion [19], parametric instability [20], and stochastic effects [21].

The $\mathcal{PT}$ symmetry-breaking threshold depends on several factors, most notably on the size of the system. Usually, this threshold decreases with the increase in the number of elements (for example, waveguides) in the system [22]. However, interesting exceptions are encountered in the discrete circular waveguide arrays, where the threshold changes in a step-like fashion with the increase in the number of waveguides [23,24]. Similar size effects were encountered in complex photonic crystals [25,26]. At the same time, longitudinal modulations of the parameters of the $\mathcal{PT}$-symmetric systems also notably affect the symmetry-breaking threshold [15]. An interesting approach to control the $\mathcal{PT}$-symmetry breaking threshold was introduced in [27], where it was shown that the geometric twist leads to the non-monotonic variation of this threshold in multi-core fibers with amplifying and absorbing cores. In the tight-binding approximation, the twist introduces Peierls phases in the coupling constants between cores of such a fiber and leads to the appearance of an artificial gauge field [27,28]. It should be mentioned that the twisted multi-core fibers provide a unique setting, where one can study the interplay between size effects and longitudinal modulations.

The aim of this Letter is to study the impact of *twisting* on the dynamics of beam propagation in $\mathcal{PT}$-symmetric multi-core fibers using a *continuous* and nonlinear model that accounts for the transformation of the field distributions within individual cores (see [29] for an example of such transformation) and radiative losses that are omitted in the tight-binding approach. We first discuss how symmetry-breaking threshold in a linear system is connected with the collision of eigenvalues of different fiber modes upon increase of the twisting rate. Then we consider a nonlinear fiber and discuss the dependence of switching length on the twisting rate, nonlinearity strength, and the strength of gain/losses. We calculate the nonlinear $\mathcal{PT}$ symmetry-breaking threshold that may be substantially lower than its linear counterpart.

We consider the propagation of light beams along the $z$-axis of the multi-core twisted fiber with alternating absorbing and amplifying cores and with the focusing cubic nonlinearity. The evolution of the dimensionless field amplitude $q$ in the fiber is governed by the nonlinear Schrödinger equation:

$$i\frac{\partial q}{\partial z} = -\frac{1}{2}\left(\frac{\partial^2 q}{\partial x^2} + \frac{\partial^2 q}{\partial y^2}\right) - R(r,z)q - \sigma|q|^2 q \quad (1)$$

Here, the longitudinal $z$ and the transverse $\mathbf{r}=\{x,y\}$ coordinates are normalized to the diffraction length $kr_0^2$ and the characteristic transverse scale $r_0$, respectively; $k$ is the wavenumber; and $\sigma \geq 0$ is the strength of the focusing cubic nonlinearity (in the linear medium one has $\sigma = 0$). The refractive index landscape $\mathcal{R}$ varies with $z$ due to the longitudinal twist of the fiber, but at any distance $z$ the symmetry $\mathcal{R}(\mathbf{r},z)=\mathcal{R}^*(-\mathbf{r},z)$ is preserved. The fiber is composed of $2N$ single-mode cores placed symmetrically on a ring, whose radius $r$ is adjusted for each $N$ in such a way that the distance $d$ between the adjacent cores remains fixed. The cores with gain and losses alternate along the ring, hence the complex refractive index profile in the $m$-th core is given by $(p_\mathrm{r}+i^{2m+1}p_\mathrm{i})e^{-[(x-x_m)^2+(y-y_m)^2]^2/w^4}$, where $w$ is the width of the core; $x_m(z)$ and $y_m(z)$ are the coordinates of the core centers on the ring; $p_\mathrm{r}$ is the real part of the refractive index and $p_\mathrm{i}$ is the strength of gain/losses. The coordinates of the core centers vary with the distance $z$ periodically, as

$$\begin{aligned} x_m(z) &= x_{m0}\cos(\alpha z) - y_{m0}\sin(\alpha z), \\ y_m(z) &= x_{m0}\sin(\alpha z) + y_{m0}\cos(\alpha z), \end{aligned} \quad (2)$$

where $\alpha$ is the rotation frequency (twisting rate). Further, we set the distance between cores $d=1.7$, the width of the individual cores $w=0.5$ and focus our attention on the impact of *nonlinearity*, rotation frequency, and gain-losses on the evolution dynamics. We consider structures possessing up to $2N=6$ cores. In contrast to discrete models, the continuous model considered here takes into account radiative losses, which grow with an increase of the rotation frequency $\alpha$, and unavoidable transformation of the modal fields due to the nonlinearity and rotation.

To illustrate the impact of the twist on the coupling strength, we first consider the simplest two-core *linear* structure with $p_\mathrm{r}=8$. At $z=0$ only one (amplifying) core was excited with the guided mode of the isolated waveguide calculated at $p_\mathrm{i}=0$. This choice of the input field allows to minimize radiative losses at the initial stages of propagation. In the conservative fiber $(p_\mathrm{i}=0)$ the energy flows $U_{1,2}(z)=\int_{\Omega_1,\Omega_2}|q|^2\,dxdy$ in the two cores (integration is performed over the half-spaces $\Omega_{1,2}$ containing cores 1 and 2) oscillate out-of-phase even at $\alpha \neq 0$, so that the total energy flow $U_1+U_2$ is conserved. In contrast, in the $\mathcal{PT}$-symmetric fiber the phase shift between $U_{1,2}(z)$ dependences appears that is different from $\pi$ and is determined by $p_\mathrm{i}$. The total energy flow $U_1+U_2$ is not conserved at $p_\mathrm{i}\neq 0$, but it oscillates periodically [2,3]. The amplitude of oscillation of the energy flows $U_{1,2}$ in two cores increases with the increase of $p_\mathrm{i}$ [see Fig. 1(a), where all $p_\mathrm{i}$ values are below symmetry breaking threshold $p_\mathrm{i}^\mathrm{cr}\approx 0.22$]. At $\alpha=0.314$, the period of oscillation (i.e. the switching length $L$) grows with $p_\mathrm{i}$, which is clearly in contrast to the behavior encountered in the one-dimensional static $\mathcal{PT}$-symmetric structures [4,18]. The increase of the rotation frequency $\alpha$ also notably slows down the switching in the $\mathcal{PT}$-symmetric linear fibers and leads to the increase of the amplitude of energy flow oscillations within both cores [Fig. 1(b)]. From the physical point of view, the twist of the fiber leads to the centrifugal and angular energy flows, so that the field distribution in the fundamental mode in each coreshifts towards the outer periphery of the core. This reduces the overlap of the modal fields in the neighboring cores and notably slows down light tunneling, because the latter is determined by this overlap. Also, the twisting introduces angular asymmetry. In the discrete model of [27], such an angular asymmetry was taken into account by introducing Peierls phases into the coupling constants, but overall reduction of their moduli due to the rotation was not taken into account. The continuous model (1) accounts for all these effects.

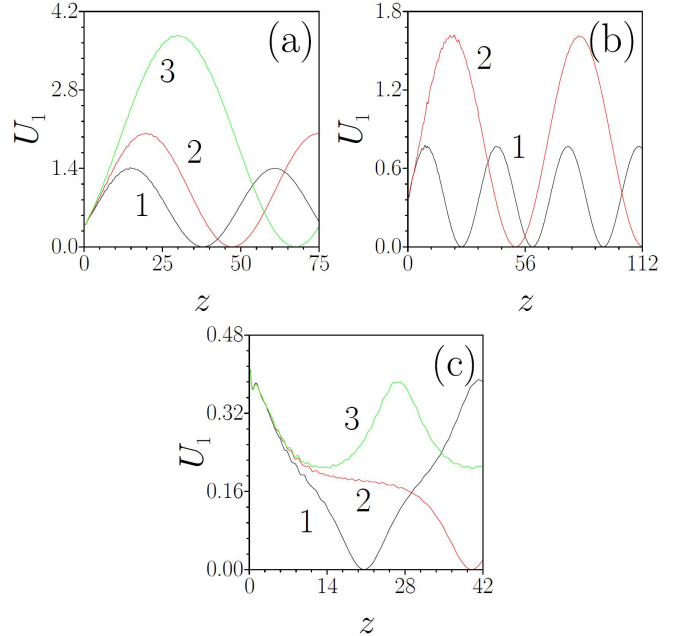

Fig. 1. (Color online) (a) $U_1(z)$ dependence at $\sigma=0$, $\alpha=0.314$ and $p_\mathrm{i}=0.19$ (1), $p_\mathrm{i}=0.2$ (2), $p_\mathrm{i}=0.21$ (3). (b) $U_1(z)$ dependence at $\sigma=0$, $p_\mathrm{i}=0.16$ and $\alpha=0.314$ (1), $\alpha=0.628$ (2). (c) $U_1(z)$ dependence at $\sigma=1$, $p_\mathrm{i}=0$ and $\alpha=0$ (1), $\alpha=0.251$ (2), $\alpha=0.283$ (3). Here $p_\mathrm{r}=8$.

To provide insight into the linear propagation dynamics in twisted fiber, we calculate its linear eigenmodes $q=w_k e^{ib_k z}$ and the corresponding eigenvalues $b_k$ in the coordinate frame that co-rotates with the fiber. The transition to this coordinate frame is described by the formulas identical to Eqs. (2). In the rotating coordinate frame, the refractive index landscape $\mathcal{R}(\mathbf{r}')$ is independent of $z$, but additional Coriolis terms appear in Eq. (1) that lead to the centrifugal energy flows. The eigenvalues of linear modes supported by four- and six-core fibers are shown in Fig. 2, as functions of the rotation frequency $\alpha$ at $p_i = 0$, $p_r = 10$. All eigenvalues monotonically grow with the increase of $\alpha$, but for illustrative purposes we removed this growth in Fig. 2 by subtracting from each eigenvalue $b_k$ the eigenvalue of the mode that was the lowest at $\alpha = 0$. One can see that the twisting leads to the *collision* of different eigenvalues. This has interesting implications for the propagation dynamics. There are only two modes in the two-core fiber (not shown in Fig. 2). When their eigenvalues approach each other, the switching length $L \sim |b_1 - b_2|^{-1}$ increases [see Fig. 1(b) and the dashed line in Fig. 5(a)], so that one can observe complete arrest of the coupling. In the four- and six-core fibers, the switching is never arrested completely, since all eigenvalues never collide at one point (still several pairs of eigenvalues can collide practically for the same $\alpha$). However, for particular $\alpha$ values the stationary field configurations appear that are built only from the modes colliding at this $\alpha$ and are characterized by strongly inhomogeneous energy distributions among the cores.

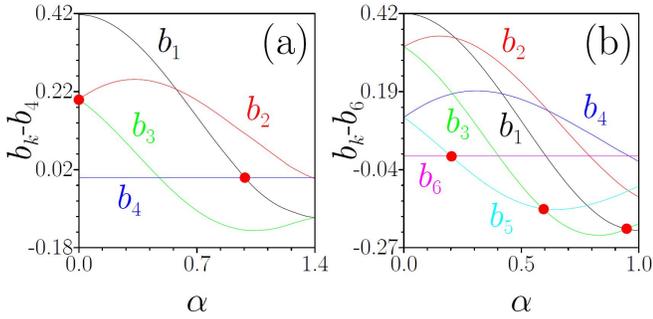

Fig. 2. (Color online) Variation of the eigenvalues $b_k$ of modes of the system with four (a) and six (b) cores with an increase of $\alpha$ at $p_i = 0$. Red dots indicate the crossing points where nonzero imaginary parts of the colliding eigenvalues appear for small nonzero $p_i$. Here $p_r = 10$.

It should be stressed that for large enough rotation frequency, the modes of the fiber become leaky and acquire small delocalized background – a signature of radiative losses. We show eigenvalues only for the rotation frequencies at which the modes are well localized and radiative losses are absent. Thus, in the two-core fiber we actually did not reach the frequency $\alpha$ at which eigenvalues $b_1$ and $b_2$ collide and a complete arrest of the coupling occurs. Still, in four- and six-core fibers, the collision of eigenvalues is clearly visible, since it occurs at smaller frequencies.

One obtains similar $b_k(\alpha)$ dependencies in the $\mathcal{PT}$-symmetric twisted fiber with $p_i \neq 0$, but now pairs of eigenvalues can collide also upon increase of the strength of gain/losses $p_i$. This collision leads to the appearance of pairs of complex-conjugate eigenvalues and it indicates the breakup of the $\mathcal{PT}$ symmetry. In this regime, some modes grow or decay exponentially. Usually, the breakup of the $\mathcal{PT}$ symmetry occurs above the critical level of gain/losses $p_i = p_i^{cr}$. However, for the rotation frequencies $\alpha$ at which the collision of one or several pairs of eigenvalues takes place already at $p_i = 0$ (they are indicated with the red dots in Fig. 2), the spectrum becomes complex even for infinitesimal $p_i$ values; that is, around these frequencies the $\mathcal{PT}$ symmetry-breaking threshold in $p_i$ vanishes. Figure 3 shows the dependence of this threshold on the rotation frequency for the fibers with two, four, and six cores. The threshold is a non-monotonic saw-tooth-like function of $\alpha$. Minima correspond to collision points for pairs of eigenvalues in the conservative case (insets schematically show eigenvalues responsible for the vanishing of the threshold). The interval between zeros of $p_i^{cr}$ decreases with the increase of the number of cores in the fiber. It should be stressed that in accordance with the results of [23] for static circular arrays, the $\mathcal{PT}$ symmetry-breaking threshold at $\alpha = 0$ is nonzero for odd $N$ and zero for even $N$. Notice also that in the continuous model, the dependence $p_i^{cr}(\alpha)$ is not periodic, in contrast to the situation encountered in the discrete setting [27]. Figure 3 shows how rotation qualitatively changes the spectrum of the $\mathcal{PT}$ system.

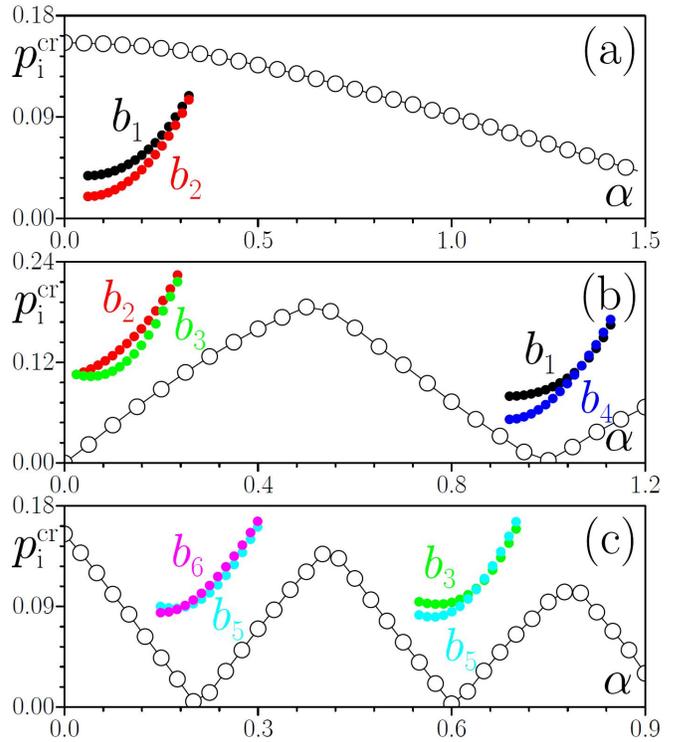

Fig. 3. (Color online) Critical gain/loss level at which $\mathcal{PT}$ symmetry-breaking occurs at $p_r = 10$, as a function of the rotation frequency $\alpha$, for a system of two (a), four (b), and six (c) waveguides. Insets indicate eigenvalues whose collision leads to symmetry breaking already at small $p_i$ values. In (a) the collision of eigenvalues $b_{1,2}$ occurs for $\alpha > 1.5$, so the inset is only to indicate how these eigenvalues approach each other.

The inclusion of the focusing nonlinearity substantially changes the switching dynamics in the twisted fiber, even at $p_i = 0$ [Fig. 1(c)]. Oscillations of the energy flow in the excited core become non-harmonic (the input mode is normalized such that its peak amplitude equals 1). The dependence $U_1(z)$ can be approximated by the Jacobi elliptic cn-function, whose period diverges as $\alpha \to \alpha_{cr}$ (the critical value of $\alpha_{cr} \approx 0.26$ at $\sigma = 1$). For $\alpha > \alpha_{cr}$ the switching becomes incomplete, and the depth of energy oscillations in the excited core decreases: the dependence $U_1(z)$ is well-approximated by the Jacobi elliptic dn-function.

Similar modification of dynamics is observed at fixed $\alpha$ upon increase of the nonlinearity strength in twisted $\mathcal{PT}$-symmetric fiber with $p_i \neq 0$. In this case, if $\alpha, p_i$ are selected such that symmetry is not broken in the linear case, the increase of focusing nonlinearity also leads to gradual increase of the switching length (see Fig. 4). In the $\mathcal{PT}$-symmetric fiber the arrest of the switching simultaneously implies a transition to the regime where the power in the amplifying core starts to grow exponentially, which leads to collapse-like dynamics, because almost no light can now couple into the absorbing core.

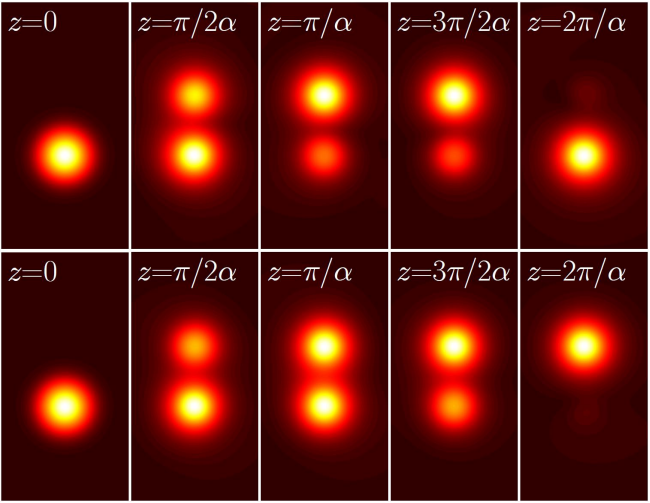

Fig. 4. (Color online) Switching dynamics in the rotating two-channel system at $p_i = 0.1$, $p_r = 8$, $\alpha = 0.251$ and $\sigma = 0$ (a), $\sigma = 0.37$ (b).

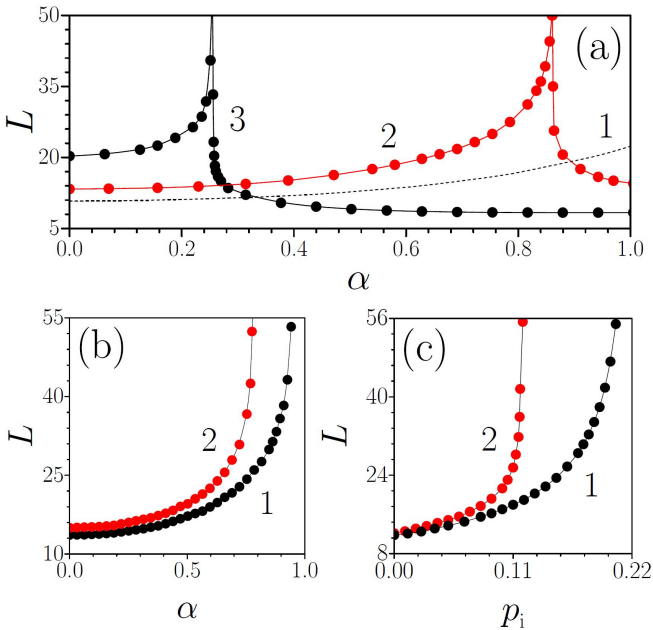

Fig. 5. (Color online) Switching length as a function of the rotation frequency (a) at $p_i = 0$ for $\sigma = 0$ (curve 1), $\sigma = 0.7$ (curve 2), $\sigma = 1$ (curve 3) and (b) at $\sigma = 0.3$ for $p_i = 0.05$ (curve 1), $p_i = 0.07$ (curve 2). (c) Switching length as a function of the strength of gain/losses $p_i$ at $\alpha = 0.314$ for $\sigma = 0$ (curve 1) and $\sigma = 0.3$ (curve 2). Here $p_r = 8$.

Figure 5 presents the central result of this Letter and illustrates how the switching length can be controlled by the rotation frequency, nonlinearity, and gain/losses. In the conservative twisted fiber, the switching length diverges at a certain rotation frequency where the complete switching (cn-type oscillations of the energy flow in individual cores) is replaced by the incomplete one (dn-type oscillations of the energy flow in the excited core). In the latter case, the switching length can be defined as a distance at which the first minimum of $U_1(z)$ is achieved. The critical rotation frequency $\alpha_{cr}$, at which switching length diverges, monotonically decreases with the increase of the nonlinearity strength $\sigma$ [Fig. 5(a)]. Importantly, similar growth and subsequent divergence of the switching length with the increase of $\alpha$ is observed in the nonlinear $\mathcal{PT}$-symmetric fiber for different levels of gain/losses [Fig. 5(b)]. However, in contrast to the conservative case, here one cannot observe the regime with incomplete switching – once energy concentrates predominantly in one core at $\alpha > \alpha_{cr}$, one observes its subsequent exponential growth. Notice that $\alpha_{cr}$ decreases with an increase of $p_i$. Finally, the switching length also strongly depends on the strength of gain/losses, as shown in Fig. 5(c). One can see from the figure that the threshold $p_i^{cr}$ below which one observes a periodic switching, substantially decreases in the presence of the focusing nonlinearity. Thus, the general conclusion can be drawn that the nonlinear threshold for $\mathcal{PT}$ symmetry breaking is always lower than the linear one. The dependence of the nonlinear threshold on the rotation frequency is similar to that of the linear threshold [Fig. 3(a)]: the breakup of symmetry at $\sigma = 0.3$ occurs at $p_i^{cr}$ values that are approximately two times lower than the linear threshold.

Summarizing, we have shown that the switching dynamics in the $\mathcal{PT}$-symmetric twisted fibers nontrivially depends on the amplitude of gain and losses, and especially on the rotation frequency. The discovered possibility of tuning the $\mathcal{PT}$ symmetry-breaking threshold by the nonlinearity and rotation frequency may be useful for light-by-light control, switching, and routing in dissipative structures.

We acknowledge support from the Severo Ochoa program (SEV-2015-0522) of the Government of Spain, from Fundacio Cellex, Generalitat de Catalunya, and CERCA. Work at the Texas A&M University at Qatar is supported by the NPRP 8-028-1-001 project of the Qatar National Research Fund (a member of the Qatar Foundation). F. Ye and X. Zhang acknowledge the support from NSFC (grant 61475101).